\documentclass[journal]{vgtc}                     

\onlineid{5895}

\vgtccategory{Research}

\vgtcpapertype{application/design study}

\PassOptionsToPackage{svgnames}{xcolor}
\usepackage{xcolor}
\usepackage{tikz}

\usepackage{braket}

\usepackage{mfirstuc}

\usepackage{gensymb}
\usepackage{soul}

\usepackage{tgheros}

\definecolor{legacyColor}{RGB}{232,232,232}
\newcommand*\legacy[1]{\tikz[baseline=(char.base)]{
            \node[shape=rectangle,fill=white, draw=black, text=black, inner sep= 2pt,minimum size=11pt,rounded corners=3pt] (char) {#1}}}

\definecolor{lineColor}{RGB}{62,50,120}
\newcommand*\quantum[1]{\tikz[baseline=(char.base)]{
            \node[shape=rectangle,fill=white, draw=lineColor, text=lineColor, inner sep= 2pt,minimum size=11pt,rounded corners=3pt] (char) {#1}}}

\definecolor{myColor}{RGB}{136,137,136}
\newcommand*\component[1]{\tikz[baseline=(char.base)]{
            \node[rectangle,draw=myColor, line width=0.5mm,rounded corners=0.7mm,text=white, fill=myColor, inner sep=0pt,minimum size=11pt](char) {\fontfamily{qhv}\selectfont{#1}}}}

\newcommand*\subcomponent[1]{\tikz[baseline=(char.base)]{
            \node[rectangle,draw=myColor, line width=0.4mm,rounded corners=0.9mm,text=myColor, fill=white, inner sep=1pt,minimum size=11pt](char) {\small \fontfamily{qhv}\selectfont{#1}}}}

\newcommand{\wy}[1]{\textcolor{brown}{[Yong: #1]}}

\newcommand{\revise}[1]{\textcolor{black}{#1}}

\newcommand{\toolName}{\textit{VIOLET}}

\title{\toolName{}: \underline{V}isual Analyt\underline{i}cs f\underline{o}r Exp\underline{l}ainable Quantum N\underline{e}ural Ne\underline{t}works}

\author{%
  \authororcid{Shaolun Ruan}{0000-0002-6163-9786},
  \authororcid{Zhiding Liang}{0000-0002-7568-0165},
  \authororcid{Qiang Guan}{0000-0002-3804-8945},
  \authororcid{Paul Griffin}{0000-0002-1656-421X},
  \authororcid{Xiaolin Wen}{0000-0002-8562-7640},
  \authororcid{Yanna Lin}{0000-0003-3730-0827}, and
  \authororcid{Yong Wang}{0000-0002-0092-0793}
}

\authorfooter{
  \item
S. Ruan, P. Griffin, X. Wen and Y. Wang are with Singapore Management
University. E-mail: slruan.2021@phdcs.smu.edu.sg, \{paulgriffin, xiaolinwen, yongwang\}@smu.edu.sg.
  \item
  	Z. Liang is with University of Notre Dame.
  	E-mail: zliang5@nd.edu.

  \item Q. Guan is with Kent State University.
  	E-mail: qguan@kent.edu.

    \item Y. Lin is with The Hong Kong University of Science and Technology.
  	E-mail: ylindg@connect.ust.hk.

   \item Y. Wang is the corresponding author.
}

\abstract{%
With the rapid development of Quantum Machine Learning, quantum neural networks (QNN) have experienced great advancement in the past few years, harnessing the advantages of quantum computing to significantly speed up classical machine learning tasks.
Despite their increasing popularity, the quantum neural network is quite counter-intuitive and difficult to understand, due to their unique quantum-specific layers (e.g., data encoding and measurement) in their architecture. It prevents QNN users and researchers
from effectively understanding its inner workings and exploring the model training status.
To fill the research gap, we propose \toolName, a novel visual analytics approach to improve the explainability of quantum neural networks. 
Guided by the design requirements distilled from the interviews with domain experts and the literature survey, we developed three visualization views: the Encoder View unveils the process of converting classical input data into quantum states, the Ansatz View reveals the temporal evolution of quantum states in the training process, and the Feature View displays the features a QNN has learned after the training process.
Two novel visual designs, i.e., satellite chart and augmented heatmap, are proposed to visually explain the variational parameters and quantum circuit measurements respectively.
We evaluate \toolName{} through two case studies and in-depth interviews with 12 domain experts. The results demonstrate the effectiveness and usability of \toolName{} in helping QNN users and developers intuitively understand and explore quantum neural networks.
}


\teaser{
  \centering
  \includegraphics[width=\linewidth]{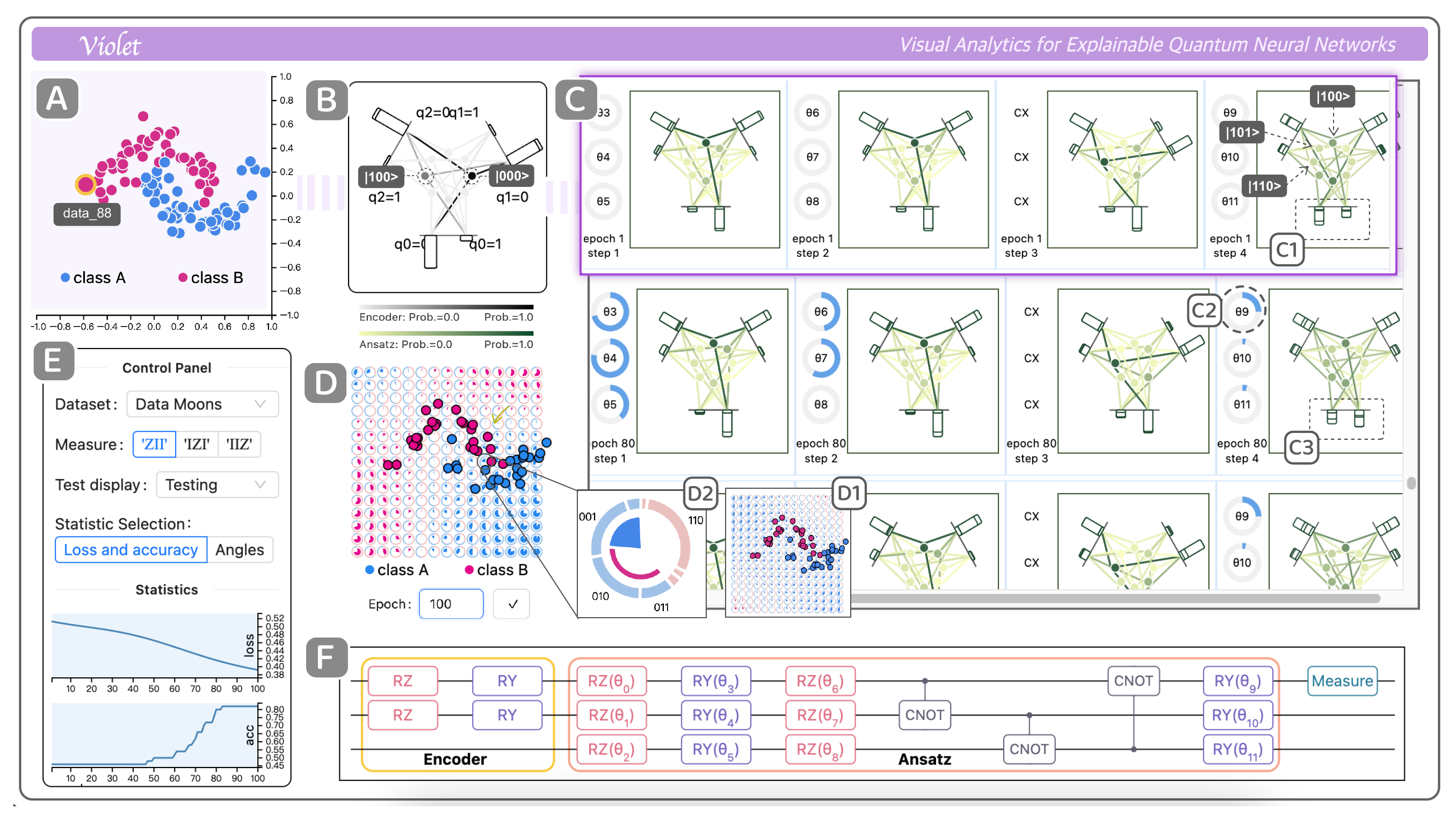}
  \caption{%
The interface of \toolName{} supports the in-depth explanation of the quantum neural networks. Encoder View (B) allows the unveiling of the inner mechanisms of the data encoding process, working with a novel design \textit{satellite chart}. Ansatz View (C) makes users aware of the training process of the ansatz in terms of the rotation angles. Feature View (D) allows the representation of what the model has learned through the color distribution of a novel design called \textit{augmented heatmap}. 
Several complementary views (A) (E) (F) are incorporated to boost the usability of \toolName{}.
}
  \label{fig:case_1}
}

\graphicspath{{figs/}{figures/}{pictures/}{images/}{./}} 

\usepackage{tabu}                      
\usepackage{booktabs}                  
\usepackage{lipsum}                    
\usepackage{mwe}

\usepackage{physics}
\usepackage{braket}

\usepackage{mathptmx}                  

\begin{document}


\firstsection{Introduction}

\maketitle

Quantum Computing (QC) has witnessed a remarkable development in recent years with the proliferation of quantum computers~\cite{preskill2018quantum, national2019quantum}. 
For example, Google conducted an experiment demonstrating quantum supremacy ~\cite{arute2019quantum} and recently shown promising results regarding quantum error correction~\cite{google2023suppressing}.
Among all the applications, machine learning is an active field of quantum computing and has led to the rapid growth of quantum machine learning (QML).
Specifically, QML is an emerging interdisciplinary research direction utilizing quantum computing to solve classical machine learning problems with a significant speed-up~\cite{zhang2020recent,schuld2019quantum,wiebe2014quantum}.
Analogous to neural networks~\cite{hubregtsen2021evaluation, maheshwari2021variational, cerezo2022challenges}, the most crucial ingredient in QML is Quantum Neural Networks (QNNs).
Quantum neural networks leverage a sequence of quantum gates to operate on quantum states with some of the quantum gates having variational parameters to be trained, making them also called \textit{variational quantum circuits}~\cite{cerezo2022challenges,cerezo2021variational,benedetti2019erratum}.
Numerous prior works have studied how to advance quantum neural networks from different perspectives~\cite{schuld2020circuit, chen2020hybrid, oh2020tutorial}.



\begin{figure}[t]
  \centering 
  \includegraphics[width=0.9\linewidth
  ]{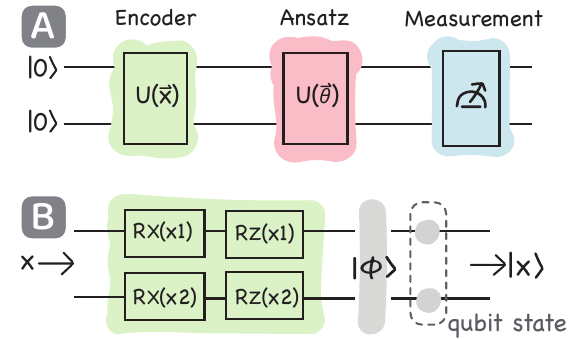}
  \caption{A two-qubit example of quantum neural networks~\cite{qnn}. (A) The architecture of the circuit, including the data encoding, ansatz, and the final measurement. (B) The detailed illustration of the data encoder and the subsequent quantum state $\ket{\phi}$, where the qubit state $\ket{\varphi}$ satisfies $\ket{\phi} = \ket{\varphi_1} \otimes \ket{\varphi_2}$.
}
  \label{fig:1}
\end{figure}

Despite the proliferation of quantum neural networks in recent years, they actually suffer from the same black-box problems as classical neural networks.
Specifically, they have proven to be counter-intuitive and notably arduous for people to understand their abstract concepts and underlying working mechanisms~\cite{heese2023explaining,daskin2023explainability}.
Note that the overall structure of quantum neural networks can be generally divided into three \textit{ quantum-specific layers}, \textit{i.e.,} data encoding, ansatz training, and circuit measurement (Figure \ref{fig:1}\component{A}), where the ansatz is the layers containing trainable parameters.
By surveying prior research on quantum neural networks~\cite{blance2021quantum, li2022quantum,zhou2023quantum,bausch2020recurrent, oh2020tutorial,tian2023recent} and working closely with six experts in quantum neural networks, 
we found that the non-transparency issues derive from these \textit{quantum-specific components}~\cite{qnn, cerezo2021variational,li2022quantum}. 
For example, it is challenging for quantum neural network users and developers to understand how a prediction is made based on the basic units of the network (\textit{i.e.,} the basis states) and how the single-qubit states manipulated by rotation gates determine those intermediate basis states.
Similarly, people also struggle to grasp how the circuit is measured from the quantum states back to the scalar values, because the principle of the measurement is intrinsically based on quantum mechanics which will prevent the users without relevant background from understanding the workings with ease. 

However, it is not a trivial task to address the above issues.
Particularly, the essential information to understand how they are working is the basis states (like symbol $\ket{\phi}$ in Figure \ref{fig:1}\component{B}), where the single-qubit states (a.k.a., qubit state, as shown in Figure \ref{fig:1}\component{B}) can be intrinsically entangled together.
It is difficult to explore the evolution of basis states and the correspondence between basis states and single-qubit states. 
Meanwhile, we already know that the measurement process converts the quantum states into the final prediction results, and people expect to achieve an intuitiveness of what ``features'' the model has learned.
But the point is how to reflect the predicted classes of data points and explain this measurement process simultaneously within the constraint of quantum mechanics remains a demanding task.
Prior studies have presented various visualization approaches to improve the explainability of deep neural networks in traditional computing~\cite{liu2017towards, choo2018visual, vilone2020explainable, mohseni2021multidisciplinary, chalkiadakis2018brief}, covering deep neural network models like convolutional neural networks~\cite{liu2016towards}, generative adversarial networks~\cite{kahng2018gan}, and graph neural networks~\cite{jin2022gnnlens,liu2022visualizing}.
Despite their effectiveness for traditional deep neural networks, they cannot be simply applied to explain quantum neural networks due to the huge difference between traditional computing and quantum computing, such as the model components, basic units (\textit{i.e.,} quantum states), and the trainable parameters (\textit{i.e.,} rotation angles) illustrated above.

To fill the research gap, we propose \toolName{}, 
a \underline{\textbf{V}}isual analyt\underline{\textbf{I}}cs approach f\underline{\textbf{O}}r exp\underline{\textbf{L}}ainable quantum n\underline{\textbf{E}}ural ne\underline{\textbf{T}}works.
We first followed a user-centered design process~\cite{munzner2009nested} to distill the design requirements for explaining quantum neural networks by working closely with six domain experts. 
Guided by the collected design requirements, we develop a visualization system  \toolName{},
which mainly consists of three visualization views: Encoder View, Ansatz View, and Feature View, which corresponds to the three major layers in the architecture of quantum neural networks.
Specifically, Encoder View intuitively explains how the classical dataset has been encoded into quantum states. Ansatz View reveals the workings of the training process along the variational circuits.
Both Encoder View and Ansatz View are built upon a novel visual design called \textit{satellite chart}, which can intuitively explain the trainable parameters by visually displaying basis states and qubit states as well as their correspondence.
Feature View facilitates the understanding of the learned features by quantum neural network and further explains the circuit measurement, where a novel design called \textit{augmented heatmap}
is proposed to achieve both goals.
We conducted case studies and in-depth interviews with 12 domain experts with carefully-designed tasks to extensively evaluate the effectiveness and usability of \toolName{}.
The results show that \toolName{} can effectively help quantum neural network users and researchers intuitively understand the behaviors of quantum neural networks.



The major contributions of this paper can be summarized as follows:

\begin{itemize}
    \item We formulate the design requirements for visualizing and understanding
    quantum neural networks by working closely with domain experts of quantum machine learning.

    \item We develop \toolName{}, a visual analytics approach to help domain users and developers easily understand the input dataset encoding, ansatz training and final output measurement of quantum neural networks. Two novel visualization designs, \textit{i.e.,} the satellite chart and augmented heatmap, are proposed.
    To the best of our knowledge, \toolName{} is the first visualization work towards achieving explainable quantum machine learning.

    \item We conduct two case studies and in-depth expert interviews to evaluate the effectiveness and usability of \toolName{}. 

\end{itemize}

To boost the impact of our paper, we make the source code of the \toolName{} system, two visual designs, and datasets publicly-available~\footnote{https://violet-source.github.io/}.

\section{Related Work}

Our work is relevant to prior research on visualization for quantum computing and visualization for explaining deep neural networks.

\subsection{Visualization for Quantum Computing}

We classify the existing visualization approaches for quantum computing into two categories based on the objectives in the paper, \textit{i.e.,} quantum circuit visualization and quantum state visualization.

\textbf{Quantum circuit visualization. }
Basically, quantum circuits can be grouped into two types, \textit{i.e.,} static quantum circuits and variational quantum circuits~\cite{griffin2021quantum}, while most of the prior work focuses on explaining the former.
According to the application scope, we further grouped the existing work into two categories, \textit{i.e.,}  \textit{generally-applicable visualization} and \textit{algorithm-specific visualization}.
Generally applicable visualization indicates those approaches that can be applied to arbitrary quantum circuits. 
For instance, Williams et al.~\cite{williams2021qcvis} leveraged a quantum simulator to compute the probability of qubit states, making the functionality of each quantum gate easier to understand.
\revise{
Different from
static quantum circuits, QuantumEyes~\cite{ruan2023quantumeyes} aims to enhance the interpretability of quantum neural networks.}
Furthermore, to address another major challenge regarding understanding quantum computing~\cite{ding2022quantum}, \textit{i.e.,} the noises hidden in quantum devices, Ruan et al.~\cite{ruan2022vacsen} proposed a visualization tool to facilitate the noise-aware execution of quantum circuits on real quantum computers.
Different from the generally-applicable visualization, algorithm-specific methods aim at specific quantum algorithms, without the generalizability like the aforementioned approach.
For example, 
Karafyllidis et al.~\cite{karafyllidis2003visualization} studied how to improve the transparency of the Quantum Fourier Transform (QFT) via the matrix-like visualization. 

While all the above methods focus on the visualizations of \textit{static} quantum circuits, they can hardly aid the explanation of \textit{variational quantum circuits}, like quantum neural networks, which contain a set of specific properties, like the trainable variational parameters and the numerous iterations for the model training. 
In contrast, our approach \toolName{} is tailored for the interpretation of the behaviors of variational circuits.

\textbf{Quantum state visualization. }
We summarized this type of related work into two groups, \textit{i.e.,} \textit{state vector-based approaches} and \textit{probability-aware approaches}, based on whether the visualization can reflect the probability of each basis state.
For the first category, the most widely-used visual representation in the quantum computing community is called \textit{Bloch Sphere}~\cite{bloch1946nuclear}. This method visualizes the pure single-qubit state using a point on the three-dimensional unit sphere.
Extended from Bloch Sphere, there was a series of variations such as the two-qubit~\cite{makela2010n} and multiple-qubit~\cite{altepeter2009multiple} quantum state visualization. 
Apart from the 3D representation, prior work also studied how to represent the quantum state on 2D shapes.
For instance, Wille et al.~\cite{wille2021visualizing} visualized the components of state vectors using a tree-like design.
\revise{In addition, Zulehner et al.~\cite{zulehner2019efficiently} implemented the decision diagram to indicate the matrix of the quantum state's state vector. Despite its intuitiveness, this method cannot interpret the probability distribution and measurement of variational quantum circuits.}
Compared to state vector-based approaches, probability-aware methods can reflect the measured probability of the basis states building on the state vector representation.
Specifically, Galambos et al.~\cite{galambos2012visualizing} used a set of rectangles to represent the multiple-qubit system with probability awareness.
Ruan et al.\cite{ruan2023venus} introduced a geometrical representation to visualize and further explain the measured probability.
\revise{Phase disk~\cite{phasedisk} visualizes the probability of State $\ket{1}$ while it cannot support the explanation of basis states' probability.}

\revise{Although some of the aforementioned methods (e.g., Bloch Sphere) can visualize the state of a single qubit,
making it possible to reflect the functionality of Pauli rotation gates,
they cannot depict the correlation between the single qubit states and basis states due to the constraint of single qubit scenario.}
On the other hand,
our proposed approach 
can visualize the impact of the rotation angles, as well as the explanation of the measurements of the quantum neural networks in terms of the basis states.

\subsection{Visualization for Explaining Deep Neural Networks
}
With the growing complexity of both data and deep neural network models, 
various visualization approaches have been developed to help us understand~\cite{ming2018rulematrix, kahng2017Activis, selvaraju2017grad, cheng2021vbridge,wang2020visual}, diagnose~\cite{ liu2016towards, wang2019atmseer, tyagi1912navigator, jin2022gnnlens, yang2022diagnosing,sun2020dfseer}, and even improve~\cite{schneider2018integrating, xiang2019interactive, yang2022diagnosing} the models.
Specifically, existing approaches can be grouped into two categories~\cite{liu2018deeptracker}: feature-oriented and evolution-oriented.

Feature-oriented visualization approach helps to understand what features the model learns during the training process and how these features affect the predictions.
For example, Grad-CAM~\cite{selvaraju2017grad} explains the CNN model by emphasizing the crucial areas in the input image that are important for predicting a given concept, showing which features in the input images are significant.
VBridge~\cite{cheng2021vbridge} goes a further step of model explanation by associating influential features with the corresponding raw data.
These methods predominantly focus on single-modal data, like images and tabular data.
Conversely, $M^{2}Lens$~\cite{wang2021m2lens}
employs a tree-like layout to visualize multimodal features, including verbal, acoustic, and visual elements, facilitating a multi-faceted exploration of such features and multi-level visual explanations on their influences.
These approaches mainly focus on the input-level features, but more and more techniques are being developed to interpret models at the neuron-level~\cite{olah2017feature, fong2018net2vec, hohman2019Summit, park2022NeuroCartograph}. 

Evolution-oriented approaches primarily concentrate on the network training process~\cite{liu2018deeptracker}. 
These methods often capture and compare multiple model snapshots at various iterations to illustrate the model's evolutionary behaviors.
Re-VACNN~\cite{chung2016re}, for instance, offers real-time visualization of each layer’s activations and changes throughout the training of CNN models.
CNNComparator~\cite{zeng2017cnncomparator} employs matrix visualization to emphasize the weight discrepancies within a layer’s filters,
facilitating side-by-side comparisons of neuron features for two model snapshots.
However, the utility of such systems is limited by the challenges in selecting and comparing relevant filters and iterations among the many options available.

The above approaches are effective for classical neural networks, but they cannot be directly used for explaining QNNs due to the inherent quantum-specific properties of QNNs, such as data encoding, ansatz training, and circuit measurement process~\cite{qnn, cerezo2021variational, li2022quantum}, which is the focus of this paper.

\section{Background}

This section introduces the background of quantum computing relevant to our study, including the basic building blocks and the quantum neural networks.

\subsection{Building Blocks in Quantum Computing}

\textbf{Qubits}, or quantum bits, are the basic units of quantum information~\cite{nielsen2010quantum}. There are two orthonormal basis states for a qubit, \textit{i.e.,} $\ket{0}$ and $\ket{1}$.
However, unlike classical computing, a qubit can be in a state that is both $\ket{0}$ and $\ket{1}$ at the same time, which can be represented as follows:

\begin{equation}
\label{equation:1}
\ket{\psi} = \alpha \ket{0} + \beta \ket{1},
\end{equation}

\noindent where $\alpha$ and $\beta$ are complex numbers representing the amplitudes of the states. This property can allow quantum computers to perform computations at an exponentially faster rate than classical computers~\cite{rieffel2011quantum}.


\textbf{Quantum gates} are the quantum version of classical logic gates, such as the blocks in Figure \ref{fig:1}\component{B}. Each quantum gate has a unique function, such as creating superpositions or performing logical operations. For example, the Hadamard gate creates superposition, the CNOT gate entangles qubits, and the Pauli-X gate acts as a quantum NOT gate.
Quantum gates are combined to form quantum circuits, which are analogous to classical circuits but operate on quantum bits. The sequence and arrangement of gates in a quantum circuit determine the outcome of the quantum algorithms.

\textbf{Quantum states} encapsulate the information encoded in qubits and describe the overall state of a quantum system~\cite{bennett2000quantum}, as shown in Figure \ref{fig:1}\component{B} the symbol $\ket{\phi}$. Quantum states are typically represented as vectors in a complex vector space. As qubits pass through quantum gates, these states evolve dynamically, reflecting the transformations applied to the qubits.
Quantum states can involve multiple qubits, and representing their collective state requires the use of the tensor product. For instance, the state of a $N$ qubit system consists of the tensor product of $N$ \textit{single-qubit states}:

\begin{equation}
\label{equation:2}
\ket{\phi} = \ket{\varphi_1} \otimes \ket{\varphi_2} \otimes \cdots \otimes \ket{\varphi_N},
\end{equation}

\noindent where $\ket{\phi}$ represents the qubit state~\cite{wilde2013quantum, keyl2002fundamentals}, as shown in Figure \ref{fig:1}\component{B}. Building upon this, any quantum state with $N$ qubits can be represented as a linear combination of $2^{N}$ \textit{basis states}, which satisfies:

\begin{equation}
\label{equation:3}
\ket{\phi} = \alpha \cdot \ket{0 \cdots 00} + \beta \cdot \ket{0 \cdots 01} + 	\cdots + \gamma \cdot \ket{1	\cdots 11}.
\end{equation}

Similar to the single qubit state, the coefficients (\textit{e.g.,} $\alpha$) are amplitudes that describe the basis state ($\ket{0 \cdots 00}$).
Meanwhile, regarding how to calculate the single-qubit state's probability concerning the basis states, we can solve it as follows:

\begin{equation}
\label{equation:5a}
Pr(q_n=\ket{x}) = \sum_{i=0}^{2^{N-1}-1}{Pr(basis\ state_{q_n=x})}, x \in \{0,1\},
\end{equation}

\noindent where $N$ represent the total number of qubits and  $n \in [1, N]$ denotes the number of qubit. For example, the probability that the Qubit 1 equals 0 for the two-qubit circuit case can be calculated as: $Pr(q_1=0) = Pr(\ket{00})+Pr(\ket{01})$.

\subsection{Quantum Neural Networks}

We introduce the three components of quantum neural networks~\cite{qnn}, \textit{i.e.,} data encoding, ansatz, and measurement.

\textbf{Data encoding}, as the first step of a quantum neural network, is used to convert the input data to quantum states, so that it can be processed by a quantum neural network as the quantum states serve as the carriers of information~\cite{altaisky2001quantum, beer2020training}.
Thus, data encoding is essential for leveraging the unique capabilities of quantum neural networks.
Angle encoding is the most popular data encoding method~\cite{piatrenka2022quantum,wang2021exploration}. 
Given the number of qubits $n$ for data encoding, we can use a circuit $S_x$ to perform the data encoding:

\begin{equation}
\label{equation:4}
S_{x}\ket{\phi} = S_{x}\ket{0}^{\otimes n} = \ket{x},
\end{equation}

\noindent where $x$ represents the data points and the initial state $\ket{\phi}=\ket{0}^{\otimes n}$~\cite{blance2021quantum}. The data encoding performs the state preparation step, followed by the trainable model called an \textit{ansatz}.

\textbf{Ansatz} is a crucial component of a quantum neural network.
Specifically, the quantum circuit $U(\pmb \theta)$ is parameterized by a set of parameters $\pmb \theta$, the rotation angles of rotation gates are the actual variational parameters to be optimized to fit the training datasets.
At each iteration of the training process, a cost function $C$ is calculated based on the measurements of the quantum circuit for a given input state $\ket{\phi_{0}}$. 
During the training process, the properties of the ansatz vary as the rotation angles, determining the function of the rotation gates, are optimized. 
The resulting quantum state $\ket{\phi_{\pmb \theta}}$ after applying the ansatz circuit is calculated as follows~\cite{jia2019quantum, wang2021exploration}:

\begin{equation}
\label{equation:5}
\ket{\phi_{\pmb{\theta}}} = U(\pmb \theta)\ket{\phi_{0}},
\end{equation}

\noindent where $U(\pmb \theta)$ is the unitary operator that represents the ansatz circuit with its parameter values $\pmb \theta$, and $\ket{\phi_{0}}$ is the initial quantum state.
The near-term quantum neural networks always use the hardware-efficient ansatz for the implementation~\cite{kandala2017hardware,huang2023near, chandarana2023digitized}, which contains a sequence of single-qubit rotation gates, \textit{e.g.,} RX gate.

\textbf{Measurement} allows quantum neural networks to obtain classical information from quantum states, thus enabling the interpretation of quantum information. 
Specifically, measuring the \textit{expectation values} of the qubits is one of the key components in the training and learning process~\cite{li2022quantum, takaki2021learning}. Here, we take a two-qubit variation quantum circuit as an example to illustrate the expectation value $E(\pmb \theta)$, which can be described as the result of the probability of $\ket{0}$ subtracted by the probability of $\ket{1}$ if we measure Pauli operator $Z$ of the first qubit:

\begin{equation}
\label{equation:6}
E(\pmb \theta) = \sum_{i \in \{0,1\}}{Pr(\ket{0i})} - \sum_{j \in \{0,1\}}{Pr(\ket{1j})}.
\end{equation}

Based on the normalization constraint~\cite{bardroff1996simulation, nielsen2010quantum}, the sum of all basis states' probabilities is always equal to 1:

\begin{equation}
\label{equation:7}
\sum_{i \in \{0,1\}, j \in \{0,1\}}{Pr(\ket{ij})} = 1
\end{equation}


\section{Informing the design}

In this section, we introduce the preliminary study and report the design requirements distilled from the study.

\subsection{Preliminary Study}

We conducted a carefully-designed preliminary study to collect the actual design requirements faced by the researchers, following the guidance of the design study methodology by Sedlmair et al.~\cite{sedlmair2012design}. We first report the participants involved in the study; then we detail the process of the study.

\textbf{Participants:}
The study involved six domain experts in quantum computing: 5 participants from educational institutions in the U.S. (\textbf{P1-5}) and one researcher from a national research laboratory in the U.S. (\textbf{P6}).
Among them, \textbf{P1-3} and \textbf{P6} study the quantum neural networks, while \textbf{P4-5} are working on the variational quantum eigensolver (VQE) algorithm.
Notably, four participants (\textbf{P1-3}, \textbf{P6}) are professors, who are more senior and are with an average of 8.6 years of research and teaching experience, while the other two (\textbf{P4-5}) are Ph.D. students working on quantum neural networks for about 3.5 years.

\textbf{Procedures:}
The entire study was divided into two stages, \textit{i.e.,} the co-design and testing stages.
We first began the co-design process by performing one-on-one, semi-structured, hour-long interviews with all participants (\textbf{P1-6}). Every participant was encouraged to describe the realistic challenges faced during the stage of learning and understanding quantum neural networks in a think-aloud manner.
Based on their feedback, we formulated the initial design requirements and proposed low-fidelity visual designs.
Next, we presented the solutions to the two Ph.D. students (\textbf{P4-5}) and asked them to test the initial designs with the in-hand datasets extracted from the training process of the ansatzes.
This process is to guarantee that our design can seamlessly fit into their routine tasks.
Then, we invited and asked them several questions regarding their suggestions and concerns upon finishing using the designs.
During the meeting, we observed and took notes on the feedback from participants.
In accordance with the summarized feedback in the second round, we further modified the designs accordingly and developed the demo system of \toolName{} with the integration of the fine-tuned visual designs.

\subsection{Design Requirements}

We compiled six design requirements and grouped them into two categories:
quantum-specific explorations (\quantum{Quantum})
and classical neural network-applicable analysis (\legacy{Classical}).

\begin{itemize}

    \item[\textbf{R1}] \quantum{Quantum} \textbf{Provide the representation of the encoded data.}
    All participants (\textbf{P1-6}) agreed that it is challenging for people to understand the data encoding process of converting classical data into quantum states.
    They hinted to us that the intuitive properties in quantum information, \textit{e.g.,} the probability of quantum states, would be helpful to connect the dots between classical and quantum information.
    \textbf{P2} also confirmed that data encoding is the primary stumbling block for beginners to learn by transferring the classical neural network to the quantum counterpart.

    \item[\textbf{R2}] \quantum{Quantum} \textbf{Explain the effects of rotation angles.}
    All participants (\textbf{P1-6}) strongly suggested that the visualization should focus on visually interpreting how the trainable parameters, \textit{i.e.,} rotation angles, determine the quantum states.
    Moreover, \textbf{P1-2} emphasized that observing how the quantum states are determined by the trainable parameters in different iterations is of great importance.

    \item[\textbf{R3}] \quantum{Quantum} \textbf{Detail the prediction along the circuit.}
    Four participants (\textbf{P1-3, P6}) expected the approach to enable the detailed analysis of how the final prediction is made through the entire circuit. In particular, \textbf{P6} commented that focusing on how a data point's quantum state will be modified across iterations is useful in understanding the ansatz.
    Moreover, \textbf{P1} also expressed the need to \textit{``display the original quantum circuit diagram for a better alignment with the visualization''}.

    \item[\textbf{R4}] \quantum{Quantum} \textbf{Support the detailed explanation of the measurement.}
    Three participants (\textbf{P2-3, P5}) emphasized the importance of explaining the measurement visually. \textit{``I really hope there exists a function to make the calculation of the measured expectation values more intuitive.''} \textbf{P2} commented.
    \textbf{P3} also mentioned that enabling the validation of the basis states based on the normalization constraints will enhance confidence in the correctness of measured values.

    \item[\textbf{R5}] \legacy{Classical} \textbf{Build an intuition about the learned features.}
    Four participants (\textbf{P2-3}, \textbf{P5-6}) encouraged us to provide a representation of the ``features'' learned by the model.
    \textbf{P5} pointed out that the heatmap-like visualization can facilitate the finding of this pattern inspired by the classical XAI methods.
    The four participants all agreed that this intuitive representation can significantly flatten the learning curves for learners with a background in classical neural networks.

    \item[\textbf{R6}] \legacy{Classical} \textbf{Show the statistics of the model training.}
    According to the feedback from three participants (\textbf{P1}, \textbf{P3-4}), it will be necessary to reflect the statistical data, such as the charts showing the changes in accuracy, loss, and trainable parameters across the iterations. 
    Additionally, \textbf{P4} also expressed the need to visualize the training dataset enabling a more flexible user interaction.

\end{itemize}

\subsection{Dataset}

Guided by the above requirements, we implemented  the models with the TorchQuantum~\cite{tq} library and then collected the following data:

\begin{itemize}
    \item \textbf{Unique properties of variational quantum circuits. }
    We extracted the basic units of the quantum neural networks, including the intermediate quantum states, variational parameters (\textit{i.e.,} rotation angles), Pauli rotation gates, and the measurements (\textit{i.e.,} expectation values). Furthermore, we calculate the probability of single-qubit states and basis states based on the above variables. 

    \item \textbf{Conventional methods applicable to classical neural networks. }
    The conventional metrics that trace the performance of the classical neural network are also suitable for evaluating quantum models.
    Specifically, we performed the validation and then extracted the performance metrics such as the training loss and accuracy across different epochs.
\end{itemize}

\section{\toolName}

\begin{figure}[t]
  \centering 
  \includegraphics[width=\linewidth
  ]{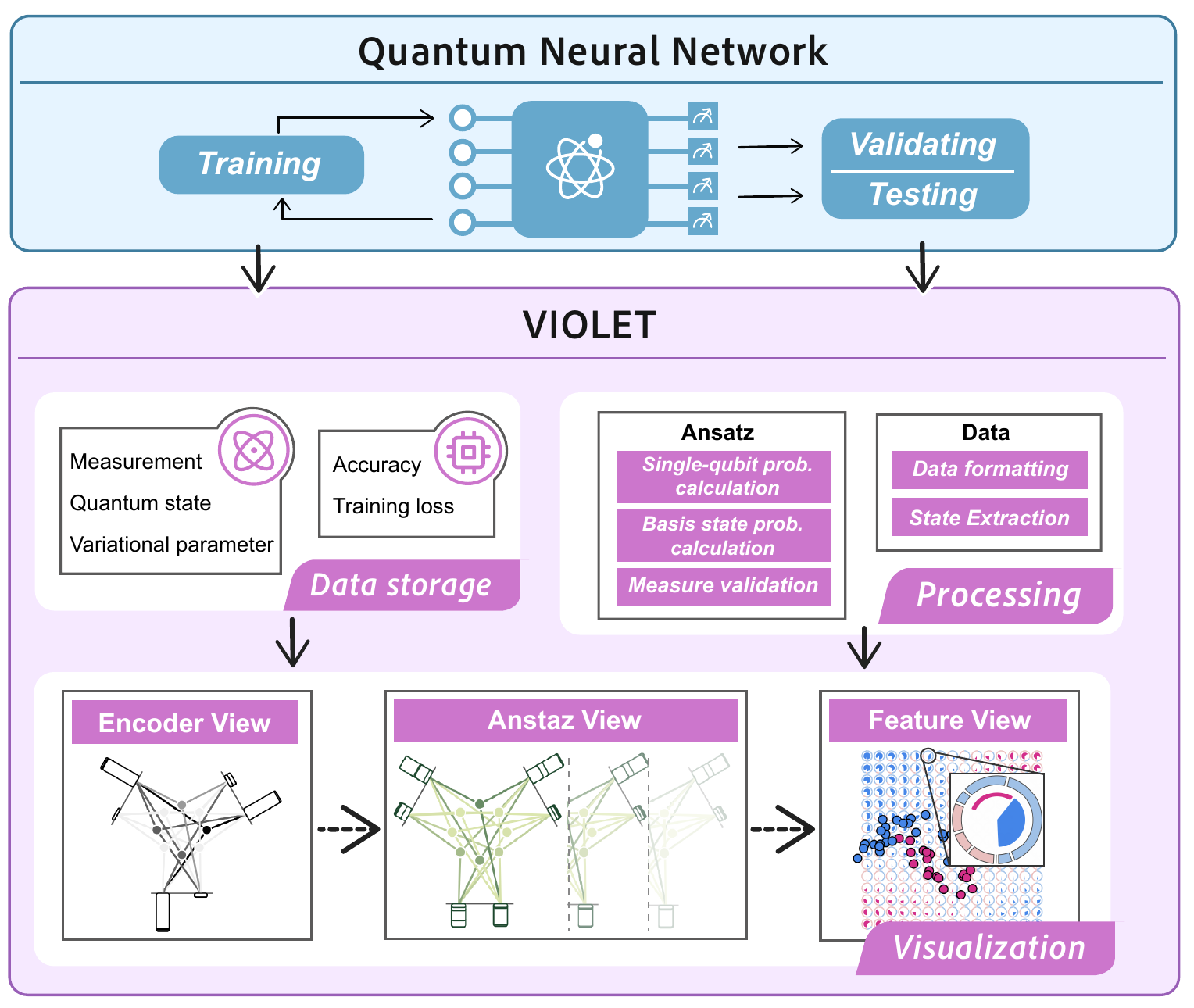}
  \caption{The system architecture of \toolName{} contains three modules (a data storage module, a processing module, and a visualization module) based on the quantum neural network. 
}
  \label{fig:2}
\end{figure}

We propose \toolName{}, a visual analytics approach to facilitate the learning and understanding of quantum neural networks.
Figure \ref{fig:2} shows the architecture of  \toolName{}.

\subsection{Encoder View}

\begin{figure}[t]
  \centering 
  \includegraphics[width=\linewidth
  ]{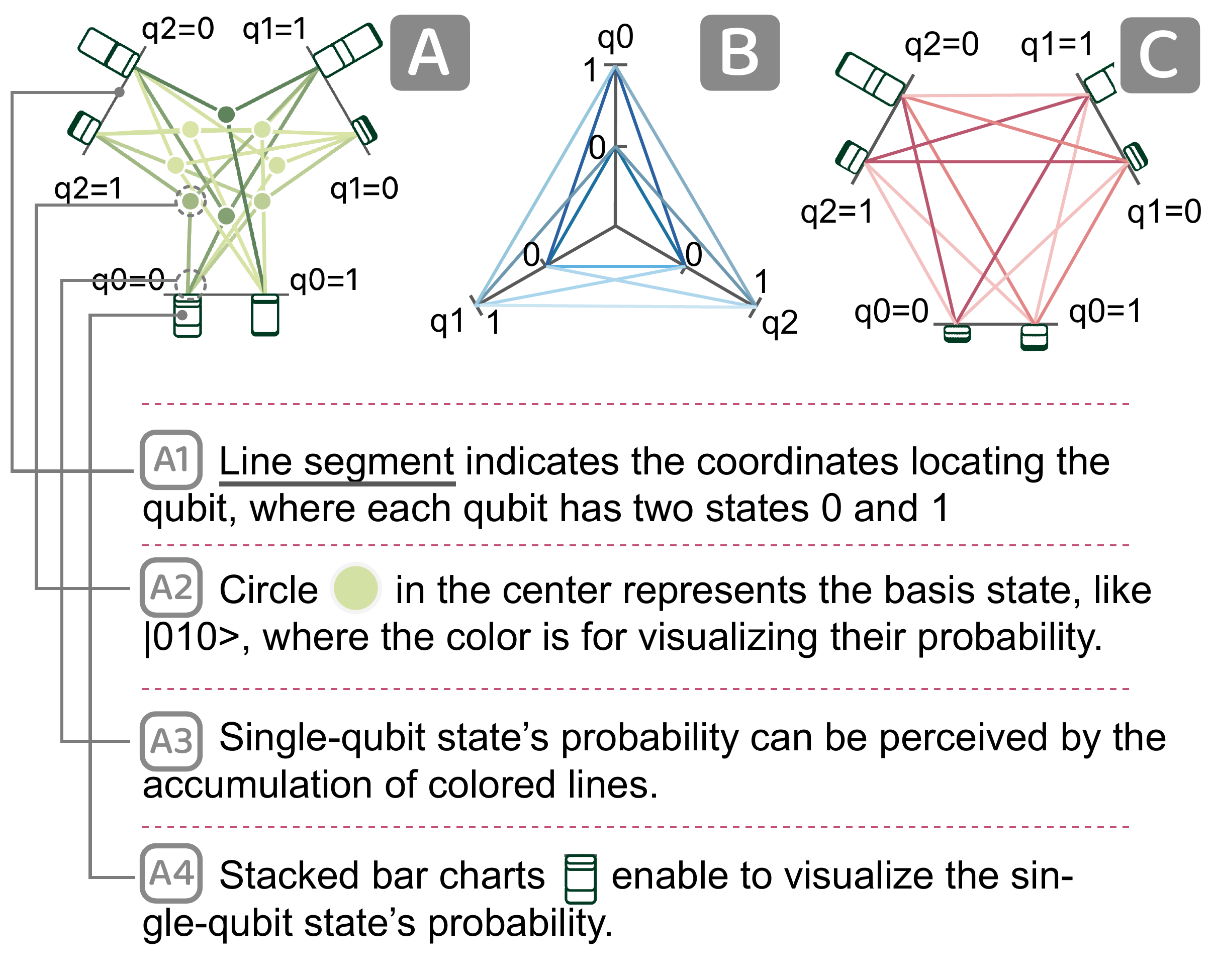}
  \caption{Satellite chart. (A) The design illustration of the satellite chart. (B) \& (C) Design alternatives of satellite chart, where (B) generates line segments with different lengths, and (C) contains severe overlapping of each line segment.}
  \label{fig:3}
\end{figure}

The Encoder View (Figure \ref{fig:case_1}\component{B}) aims to allow users to explore and understand how the data encoding step works via the visualization of the encoded classical data (\textbf{R1}), \textit{i.e.,} quantum states.
Given that the rotation gates inside the data encoder calculating the basis states are actually acting on the single qubit,
we propose a satellite chart (Figure \ref{fig:3}\component{A}), which can enable the explanation of the data encoder by correlating the single-qubit state with the basis states.

\textbf{Satellite chart.}
In this section, we use a 3-qubit quantum neural network as an example to illustrate the visual designs.
As shown in Figure \ref{fig:3}\subcomponent{A1}, three axes are used to locate the digit of the quantum states, where the axis is spaced with an equal degree from its neighboring axis.
For each digit, given there are two states for one digit position (\textit{i.e.,} $\ket{0}$ and $\ket{1}$),  we use the same position on each axis to locate the two states, \textit{e.g.,} for qubit 0, left-hand side position is for state $\ket{0}$ and the state $\ket{1}$ is on the right-hand side.
For the representation of basis states, a set of circles is arranged concentrically, forming a larger circular pattern (Figure \ref{fig:3}\subcomponent{A2}), \textit{e.g.,} eight basis states for the 3-qubit case ($2^{3}=8$).
Meanwhile, we encode the probability of each basis state as the color of the circles, ranging from \revise{light green} to dark green.
To correlate the probability of the single-qubit state with the basis state, we first utilize the lines linking from the circle of the basis state to the position where the single-qubit state is located on the respective axis.
For example, the data entity in Figure \ref{fig:3}\component{A} indicates the basis states of $\ket{110}$.
Next, we color the connection line using the color the same as the circle.
In this way, the line segments with the same digits of the basis states are gathered around the position of the axis.
According to quantum mechanics, the probability of the single-qubit values equals the sum of the probabilities of the basis state with that value, \textit{e.g.,} $Pr(q_0=1) = Pr(\ket{10})+Pr(\ket{11})$ for the two-qubit case.
In a satellite chart, the above sum is encoded by the color around the gathered area of the digit (Figure \ref{fig:3}\subcomponent{A3}).
Note that we did not consider the encoding of line width for visualizing the basis state's probability since we found that it would introduce a severe overlapping between lines, making the design fail to explain the single-qubit state's probability.
Moreover, to improve the quantity representation of the above constraint, we implement a stacked bar chart to enable a more accurate perception, where each section indicates the basis state's probability and the total height depicts the single-qubit state's probability ((Figure \ref{fig:3}\subcomponent{A4})).
We name the design ``satellite chart'' due to its notable similarity to the metaphor of a set of satellites (single-qubit states) positioning around a planet (basis states).

\textbf{Design alternatives.}
We considered several design alternatives before finalizing the current visual design. 
Figure \ref{fig:3}\component{B} uses the axes forming the star to encode the positions of the single-qubit states, while each basis state's probability is also represented by the color of the respective line segment. 
But this method encodes the states with the same semantics, \textit{e.g.,} $\ket{00}$ and $\ket{11}$, by the lines with different lengths, introducing the perceptive bias for the comparison.
Hence, we propose another design (Figure \ref{fig:3}\component{C}) to aid this issue by rotating the direction of all axes.
However, based on the experiment of the real data, we found that there are always two overlapped lines connecting the same two points of the digits, preventing one of the lines from being seen at the time.
At last, we propose the current design where line segments can be naturally separated from each other, making the perception more accurate without the line overlapping.

\subsection{Ansatz View}

The Ansatz View supports the in-depth analysis of the evolution of the ansatz and training process. 
We propose a matrix-like layout to denote the changes of ansatzes during the training process (Figure \ref{fig:case_1}\component{C}).
Since the ansatz layer uses the same type of rotation gates, \textit{e.g.,} hardware-efficient gates like RX gate, we apply the satellite chart to portray the behavior of the ansatz similar to the data encoder.

\textbf{Evolution of rotation gates (R2).}
In \toolName{}, we use the quantum state visualization to represent the evolution of the Pauli rotation gates along the training process.
We first define the quantum gates with the same implementation position as a ``step''.
To enable the exploration in terms of the quantum gates, we use the column of the matrix for the arrangement of all satellite charts of the same gate in the circuit.
As shown in Figure \ref{fig:case_1}\component{C}, the matrix cells can be unfolded if the user clicks the columns of quantum gates of interest.
For each cell of the matrix, we then use the donut chart to depict the change of the trainable parameters (\textit{i.e.,} angles of the Pauli rotation gates).
The section size of the donut chart outlines the difference in the angle change between the current epoch and the first epoch.
Note that we grouped the controlled logic gates between two groups of Pauli rotation gates as one step due to the absence of the trainable parameters for these controlled gates.

\textbf{Drill-down of the prediction process (R3).}
Apart from the analysis of the rotation gates, revealing how the quantum state of a data point is calculated and finally predicted is also crucial for the interpretability enhancement of the variational circuits.
Building on the matrix layout, we design each row of the matrix to show the prediction process of a data point.
Specifically, each cell of the row contains a satellite chart to denote the states after a step of the entire variational circuit.

\subsection{Feature View}

The Feature View aims to provide the user with an intuitiveness of what ``feature'' the quantum circuit learned (\textbf{R5}).
Meanwhile, the explanation of the measurements is also helpful for users to grasp how the measuring components work (\textbf{R4}).
To meet the aforementioned needs, we develop the Feature View with a novel design named ``augmented heatmap'' embedded into the view.

\textbf{Feature sampling.}
We first uniformly sample the feature data of each dimension; and then feed the dataset into the quantum neural network to calculate their prediction during the valid process.
Specifically, we first normalized each dimension of the features to the range $[-1, 1]$. Then, we generate a set of artificial datasets by normally sampling the values and slicing the data of each dimension into 15 pieces. This operation can fulfill the entire 2D plane and also keep a high resolution of each augmented heatmap's units, enabling users to perceive the feature's learned pattern easily.

\begin{figure}[t]
  \centering 
  \includegraphics[width=0.95\linewidth
  ]{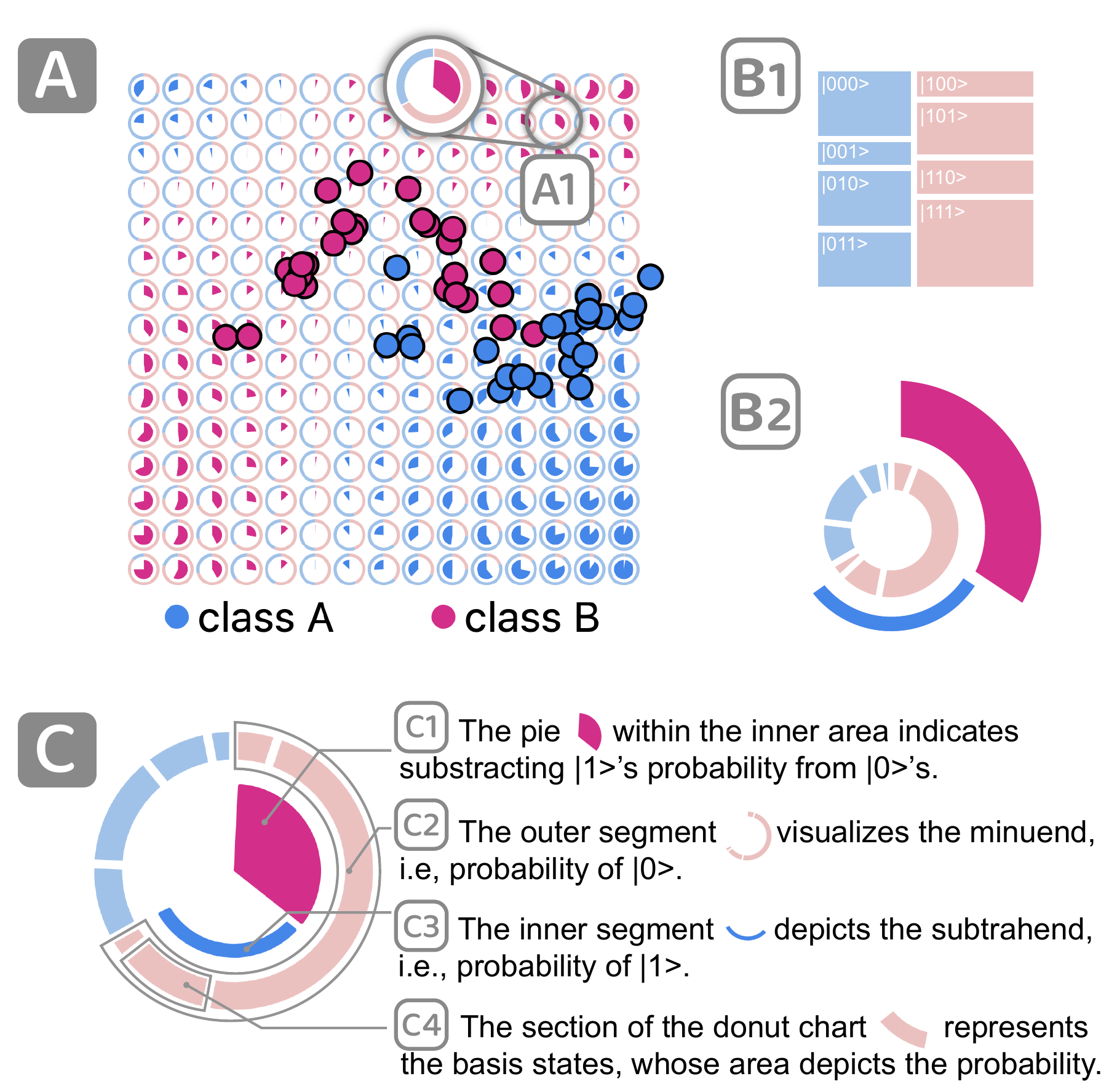}
  \caption{Feature View and augmented heatmap. (A) Feature View visualizes what the model learned via the background of a matrix consisting of augmented heatmaps. (B) The design alternatives of the augmented heatmap. (C) The fine-grained mode of the augmented heatmap, allowing the explanation of the measured values.
}
  \label{fig:4}
\end{figure}

\textbf{Augmented heatmap.}
To reveal the measurement rules (Equation \ref{equation:6}) while displaying the normalization constraint (Equation \ref{equation:7}) which allows the validation of the basis states, we propose a visual design of the augmented heatmap to solve the above challenges.
With the sampled data and their prediction above, we first apply the normalization constraint (Equation \ref{equation:7}) by encoding the probability of all basis states by the section size of the donut chart, as shown in Figure \ref{fig:4}\subcomponent{C4}.
The donut sections in light blue represent the probability of $\ket{0}$ while the donut sections in light red depict the probability of $\ket{1}$ based on Equation \ref{equation:5a}. 
Then we visualize the measured expectation values, as Equation \ref{equation:6} shows, using the section of the pie within the outer donut chart (Figure \ref{fig:4}\subcomponent{C1}).
To visually reflect what Equation \ref{equation:6} specifies, the outer larger section of the donut section is used to show the minuend (Figure \ref{fig:4}\subcomponent{C2}), while the inner arc denotes the subtrahend (Figure \ref{fig:4}\subcomponent{C3}).
In this way, the size of the pie section can be visually explained via the difference.

\textbf{Display modes. }
The augmented heatmap aims to reflect the learned ``feature'' and the explanation of the expectation value measurement simultaneously. 
To this end, the Feature View applies the coarse-grained mode of the augmented heatmap (Figure \ref{fig:4}\subcomponent{A1}) to directly reflect the confidence of the predicted class of the sampled units, which can intuitively exhibit the learned features of the ansatz.
Meanwhile, when the user clicks a coarse-grained augmented heatmap, a fine-grained mode chart (Figure \ref{fig:4}\component{C}) will pop up to visually explain how the expectation value is calculated with a more detailed illustration.

\textbf{Design alternatives. }
We considered other two design alternatives for the explanation of the measured expectation values.
Figure \ref{fig:4}\subcomponent{B1} visualizes the probability of the basis states, preserving the nature that the left-side group of blocks with light blue represents the probability of $\ket{0}$ while the block group on the other side visualizes the probability of $\ket{1}$. In this way, users can perceive which group has a larger probability via the area.
However, this method cannot visualize the prediction of the class directly without a representation of difference (e.g., Figure \ref{fig:4}\subcomponent{C1}), making the Feature View difficult to reveal the model's learned features.
Figure \ref{fig:4}\subcomponent{B2} is the initial design of the augmented heatmap.
However, this approach cannot obviously reflect the feature distribution since the outer donut section is relatively small, especially in the case where the probability of the two classes is close.
To aid all the above issues, we finalize the design as Figure \ref{fig:4}\component{C}.

\subsection{Complementary Views}

To better trace the training process and the model performance (\textbf{R6}), we embed a series of classical statistical charts into \toolName.
For example, as shown in Figure \ref{fig:case_1}\component{A}, a dataset overview is used to show the distribution of the training set, where the x- and y-axis represent the two dimensions of data points.
We then add the charts of loss and accuracy to reflect the model performance in a conventional style (Figure \ref{fig:case_1}\component{E}).
Also, the chart showing the rotation angles' changes will be rendered if the user clicks the ``angle'' button in the ``Statistic selection'' item in the control panel.
During the interview with the participants, they also found it hard to perform the comparison with the original quantum neural network.
To solve the issue, we implement a quantum circuit diagram to allow the alignment with the Ansatz View with ease (Figure \ref{fig:case_1}\component{F}), where the type of rotation gates and parameter names are explicitly highlighted.

\section{Case Study}

In this section, we report two case studies with two different quantum neural networks to demonstrate the effectiveness of \toolName{}.
To achieve this, we focus on the variational quantum classifiers since it is widely used to solve classification tasks in the near-term era~\cite{maheshwari2021variational, li2022recent, huang2021variational}. 
The users involved in our case study are two experts in quantum computing (E12 and E2) who participated in the expert interview as well.
To assess the potential usage of \toolName{}, we propose two types of workflow, \textit{i.e.,} forward exploration and backward exploration, which are used for the diagnosis of the model and debugging for the cause of incorrect prediction reversely.

\subsection{Case Study \uppercase\expandafter{\romannumeral 1} - Forward Exploration}

We worked with E12, who is a Ph.D. student with 3.5 years of experience in quantum neural networks, to explore the model's prediction process.
As suggested by E12, we reproduced the standard 2-dimensional datasets introduced by Zhou et al.~\cite{zhou2023quantum} in advance.
We then implemented the 3-qubit encoding and ansatz layer following the tutorial by Paddle Quantum~\cite{paddle}. 
He was asked to conduct the analysis task of forward exploration.

\textbf{Enhance the transparency of data encoding. }
After a glance at the scatter plot of the dataset (Figure \ref{fig:case_1}\component{B}), E12 first randomly selected a data point \textit{data\_88}, whose dimensional data is $[-0.58, 0.10]$.
By clicking the data point, the Encoding View appeared along with the satellite chart.
E12 then quickly found that among all basis states generated from the \textit{data\_88}, there are only two obvious basis states: State $\ket{000}$ in dark black with the probability of 90.5\% and State $\ket{100}$ in grey with the probability of 9.3\%.
Keeping this in mind, E12 then notices that the single-qubit probability of each qubit contains a small portion of $\ket{1}$, while the most part of the probability is $\ket{0}$, as the height of stacked bar charts indicated.
E12 commented \textit{``This is exactly what I expected. If the encoded data are all 0, the basis state will be $\ket{000}$ in the initial stage. However, after being rotated by the rotation gates of data encoding, each single-qubit state is slightly modified by the dimensional data $[-0.58, 0.10]$. The first attribute, (\textit{i.e.,} -0.58) rotates the state of the first qubit, making the circle color of $\ket{100}$ grey instead of white.''}
We reminded him of the quantum circuit diagram (Figure \ref{fig:case_1}\component{F}) showing the architecture of the state preparation process. Based on this, E12 further reported that the absence of the rotation gates on Qubit 2 explains why the probabilities of all basis states with the Qubit 2 of $\ket{1}$ are 0. This finding can also confirm the pattern that the line segment color gathered around the position of ``q2=1'' is totally white.
\textit{``I'm surprised that Quantum satellite \revise{chart} tells me how the classical data is encoded and explained by the single-qubit state.''} E12 praised.

\textbf{Identify the hidden reasons for the model prediction. }
After exploring the data encoding stage, E12 proceeded to analyze the trainable model, \textit{i.e.,} ansatz.
By glancing at the line charts in Figure \ref{fig:case_1}\subcomponent{E}, E12 noticed that the accuracy suddenly increased around Epoch 40 and stay still after Epoch 80.
E12 thus browsed the satellite charts between Epoch 40 and 80 in the Ansatz View.
Hinted by the circuit diagram, E12 noticed that the expectation value of the variational circuit is calculated by measuring Qubit 0.
Bearing this in mind, E12 easily found that the probability that Qubit 0 equals 0 is the larger one, while the probability that Qubit 0 equals 1 became larger 40 iterations later.
\textit{``This indicates that the predicted class of the data point changes from A to B, which is caused by the increasing of the probability of Basis state $\ket{100}$, $\ket{101}$ and $\ket{110}$ (indicated by their line segment color of Figure \ref{fig:case_1}\subcomponent{C1}).''}
To figure out how the rotation angles contribute to this prediction, E12 analyzed the Ansatz View and then reported that the ring (donut chart) of $\theta9$ shows a large change throughout the training process. This will increase the probability of $q0=0$ significantly increase (because the $\theta9$ acts on Qubit 0), which exactly explains why the probability of the above three basis states is much larger than before (because the first digits of the basis states are $\ket{1}$). 
E12 commented, \textit{``\toolName{} reveals the reasons that lead to the final predictions of the data points. I was really confused before using \toolName{}''}.

\textbf{Unearth the evidence of circuit measurements. }
Upon analyzing the training process, E12 then started to focus on the validation stage via the Feature View.
First, E12 inspected the initial patterns through Feature View, where the background of the entire map is color blue (Figure \ref{fig:case_1}\subcomponent{D1}).
Meanwhile, E12 noticed that the ``confidence'' of each augmented heatmap is high, indicated by the size of the pie section.
Next, E12 moved on to explore the Feature View in Epoch 100 when the model is already converged. He quickly found that the area highlighted has changed into red impacted by the corresponding set of data points with class B.
Also, E12 noticed the area where most data points are with the same class has higher confidence compared to those points with mixed classes.
And then, E12 found the data point highlighted by the yellow arrow was incorrectly classified. He thus clicked the corresponding augmented heatmap to figure out the reason. 
As shown in Figure \ref{fig:case_1}\subcomponent{D2}, the probability of $\ket{0}$  is significantly larger than the probability of $\ket{1}$, indicated by the donut sections in blue and red respectively, which reveals the rationale of this incorrect classification.
\textit{``The expected larger size of the donut section of $\ket{1}$ is mainly caused by the Basis state $\ket{110}$, offering me the clue to solve this issue.''
}


\begin{figure*}[t]
  \centering 
  \includegraphics[width=0.95\linewidth
  ]{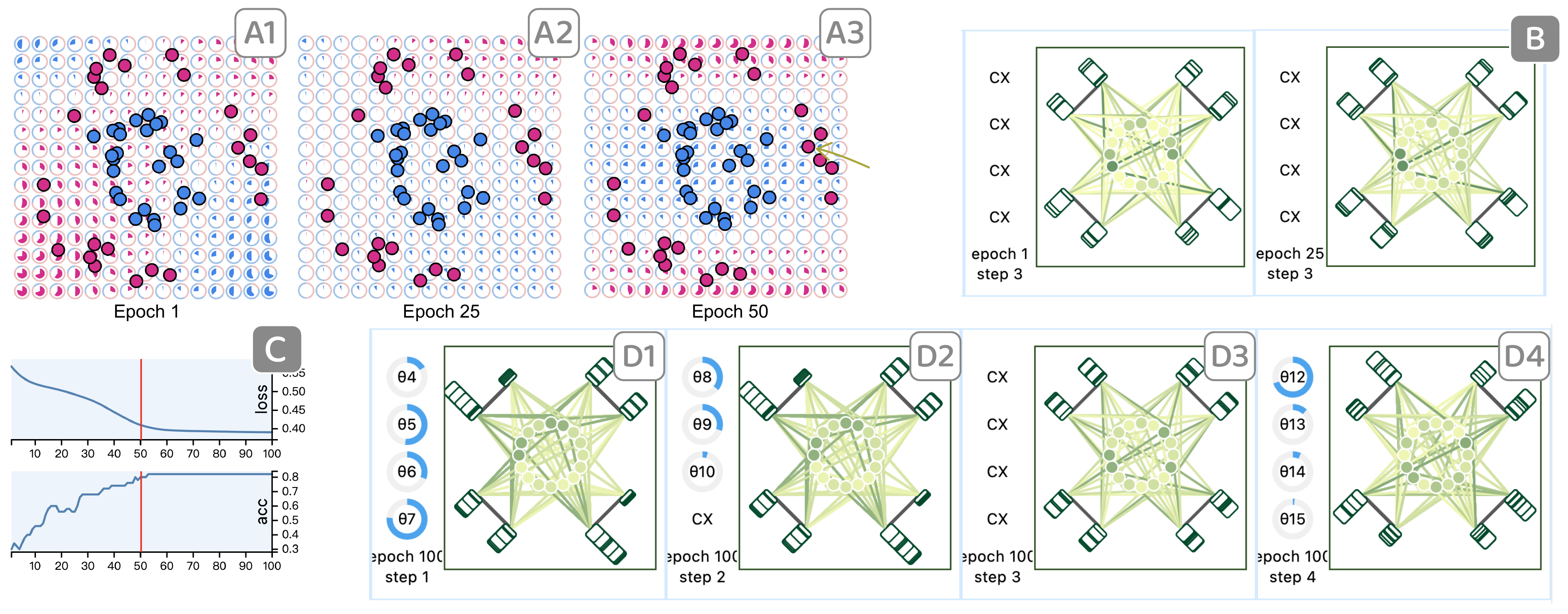}
  \caption{The case for the backward exploration. (A) Augmented heatmaps in Epochs 1, 25, and 50. (B) The satellite charts after CNOT gates in Epochs 1 and 25. (C) Statistic charts. (D) The satellite charts along the variational circuits in Epoch 100.}
  \label{fig:case2}
\end{figure*}

\subsection{Case Study \uppercase\expandafter{\romannumeral 2} - Backward Exploration}

We worked with E2, who has 8 years of working experience in quantum computing, to investigate the 4-qubit variational circuits with the dataset generated by the \textit{make\_circles} function from the \textit{scikit-learn} library.
The architecture of the circuit remains the same as in the previous case.
He was asked to use \toolName{} to conduct
the backward exploration for diagnosing quantum neural networks.

\textbf{Identify the incorrectly-classified data points. }
E2 started by observing the augmented heatmap to identify the target data point.
E2 first glanced at the line charts (Figure \ref{fig:case2}\component{C}) and easily found that the accuracy started to increase in Epoch 1 and converged around Epoch 50.
So he selected Epochs 1, 25, and 50 to analyze further.
As shown in Figure \ref{fig:case2}\subcomponent{A1}, the variational parameters were just initialized and the features learned by the model were randomly distributed, as indicated by the background augmented heatmap.
After 25 iterations of training, the uncertainty of the model is high, where the confidence of predictions for Class A and B are relatively close (Figure \ref{fig:case2}\subcomponent{A2}).
\textit{``I think the prediction will be totally random because the model seems to be uncertain for any input attributes.''}
Then, in Epoch 50, the background is divided into three layers, while the middle layers in blue are dominated by the blue dots in the center and cause the red dots to be predicted incorrectly (Figure \ref{fig:case2}\subcomponent{A3}).
\textit{``This somehow shows that the model is too weak to capture the underlying patterns in the data.''}
Thus, E2 then planned to investigate the possible ways to improve the model to be better.

\textbf{Formulate a strategy to improve the model. }
To analyze the incorrectly-classified data point highlighted by the yellow arrow,
E2 first scrolled the satellite charts to Epoch 100 and noticed that the changes of the rotation angles were becoming fewer along with decrease of the circuit depth.
as shown in Figure \ref{fig:case2}\subcomponent{D1} to Figure \ref{fig:case2}\subcomponent{D4}.
\textit{``This really makes sense to me. Closer to the measurement of the first qubit, all parameters (except CNOT gates without parameters) will `draw together' to it and other parameters will be that sensitive like in the beginning.''}
Next, after browsing all satellite charts, E2 reported that the effect of the layers of the CNOT gates is to \textit{``make all qubit states with the similar probabilities like the height of the stacked bar chart shows} (Figure \ref{fig:case2}\component{B} and Figure \ref{fig:case2}\subcomponent{D3}).''
However, the difference is that the probabilities of basis states consisting of the single-qubit states vary, which aims to \textit{``prepare the states to be classified in the final step''}.
E2 then found that the probability of $q0=0$ is higher than that of  $q0=1$, which should be exactly the opposite case according to the data point's ground-truth label, \textit{i.e.,} Class B.
This scenario indicates that the Rotation angle $\theta12$ rotates the states of Qubit 0, but does not work because $\theta12$ also needs to fit other data points to make the classification correct.
\textit{``The finding hints me the reason for the mis-classified data entities. The model performance can be improved by appending another rotation gate on Qubit 0 to make the features of those data points on edge be fitted and also make the ansatz more resilient and robust.''}

\section{Expert Interview}

To further assess the effectiveness of \toolName{}, we conducted in-depth expert interviews with target users studying quantum neural networks.

\subsection{Study Design}
\textbf{Participants and Apparatus. }
We recruited 12 participants (\textbf{E1-12}) (2 females, $age_{mean}=33.2, age_{sd}=5.2$) from six educational institutions in the U.S.
All participants were selected based on their relevance to our research topic.
Specifically, \textbf{E1-3} study Error Mitigation in Quantum Machine Learning (QML); \textbf{E4-6} are working on Quantum Uncertainty Analysis of QML; \textbf{E7-9}'s research directions are Quantum Simulator; and other three experts (\textbf{E10-12}) mainly focus on Quantum Information.
None of the above experts participated in the study to collect design requirements.
The study was conducted via the Zoom meeting based on the online interface we deployed in advance.
All participants were suggested to use a monitor with a $1920 \times 1080$ resolution.

\textbf{Tasks. }
The primary four tasks were crafted to evaluate the effectiveness of \toolName{} regarding the explanation enhancement of quantum neural networks.
To this end, we designed seven tasks and then asked all participants to accomplish them within a fixed time.
A comprehensive list of tasks is shown in Table \ref{table:1}.
Specifically, T1-4 aim to assess the participants' performance for the model's training process with the help of satellite charts, where T1-2 are presented to test the data encoding and T3-4 are for the steps of the execution procedure of the ansatz. Furthermore, T5-7 are used to evaluate the usage of the augmented heatmap, which is targeted for the testing process.

\begin{table}[t]
\centering
\begin{tabular}{c|p{0.8\columnwidth}}
\hline
T1 & Find the rationale of how classical data is encoded into the quantum states.                                                           \\ \cline{2-2} 
T2 & Identify all the basis states and then tell how they form the single-qubit states.                                                     \\ \hline
T3 & Trace back how the output quantum states were modified by the rotation gates along the quantum neural network.                    \\ \cline{2-2} 
T4 & Identify how the variational parameters (\textit{i.e.,} rotation angles) were trained and further tell their impact on the same rotation gates. \\ \hline
T5 & Compare the distribution of augmented heatmaps and then tell the model evolution.                                               \\ \cline{2-2} 
T6 & Pinpoint the impact of the training dataset on the featured learned by the model.                                                      \\ \cline{2-2} 
T7 & Tell how the incorrectly classified data point was measured based on the expectation values.                                           \\ \hline
\end{tabular}
\caption{%
All tasks are grouped by the steps along the training of quantum neural networks, \textit{i.e.,} the data encoding (T1-2), training of ansatz (T3-4), and testing stage (T5-7).
}
\label{table:1}
\end{table}

\textbf{Procedures. }
The study was undertaken using the online platform \toolName{}. We executed a one-on-one, semi-structured interview with all expert participants. 
Initially, we showcased the visual design of \toolName{}, including the illustration of a satellite chart and augmented heatmap.
We then asked all participants to complete the seven tasks on the platform.
During the process, we recorded their comments and the interactions.
The duration of the above introduction lasts about 20 minutes.
Subsequently, participants were requested to verbally describe the findings during the task procedure. This step took approximately 40 minutes. 
After the description, we encouraged participants to share their thoughts on the proposed visual designs in a think-aloud manner. 
Moreover, participants are also asked to rate \toolName{} on a 7-point Likert scale based on the post-study questionnaire (Table \ref{table:2}), focusing on design requirements we had previously outlined. The post-study interview lasted about 20 minutes.

\begin{table}[t]
\centering
\begin{tabular}{c|p{0.8\columnwidth}}
\hline
Q1  & The system can help me enhance the interpretability of quantum neural networks.                                                    \\ \cline{2-2} 
Q2  & The system shows a holistic picture of how the classical data is encoded into quantum states.                                          \\ \cline{2-2} 
Q3  & The system enables the analysis of how a data point is classified through a series of rotation gates.                                   \\ \cline{2-2} 
Q4  & The system provides me with an intuitive representation of the features learned by the model.                                                \\ \hline
Q5  & The visual design is easy to understand.                                                                                                \\ \cline{2-2} 
Q6  & The satellite chart can show how the rotation angles affect the prediction.                                                    \\ \cline{2-2} 
Q7  & The augmented heatmap can effectively explain the measured expectation values.                                                   \\ \hline
Q8  & The interaction is smooth.                                                                                                              \\ \cline{2-2} 
Q9  & The interaction can support the forward exploration of explaining the prediction and the backward exploration of the circuit debugging. \\ \hline
Q10 & The system is easy to use.                                                                                                              \\ \cline{2-2} 
Q11 & The visual designs would fit into my routine tasks of the development of quantum neural networks.                                       \\ \cline{2-2} 
Q12 & I will recommend the visual analytics system to my colleagues working on quantum computing.                                             \\ \hline
\end{tabular}
\caption{%
The questionnaire consists of four parts: the effectiveness of interpretability enhancement (Q1-4), the visual design (Q5-7), the user interactions (Q8-9), and the usability (Q10-12).
}
\label{table:2}
\end{table}

\subsection{Results}

\begin{figure}[t]
  \centering 
  \includegraphics[width=0.95\linewidth
  ]{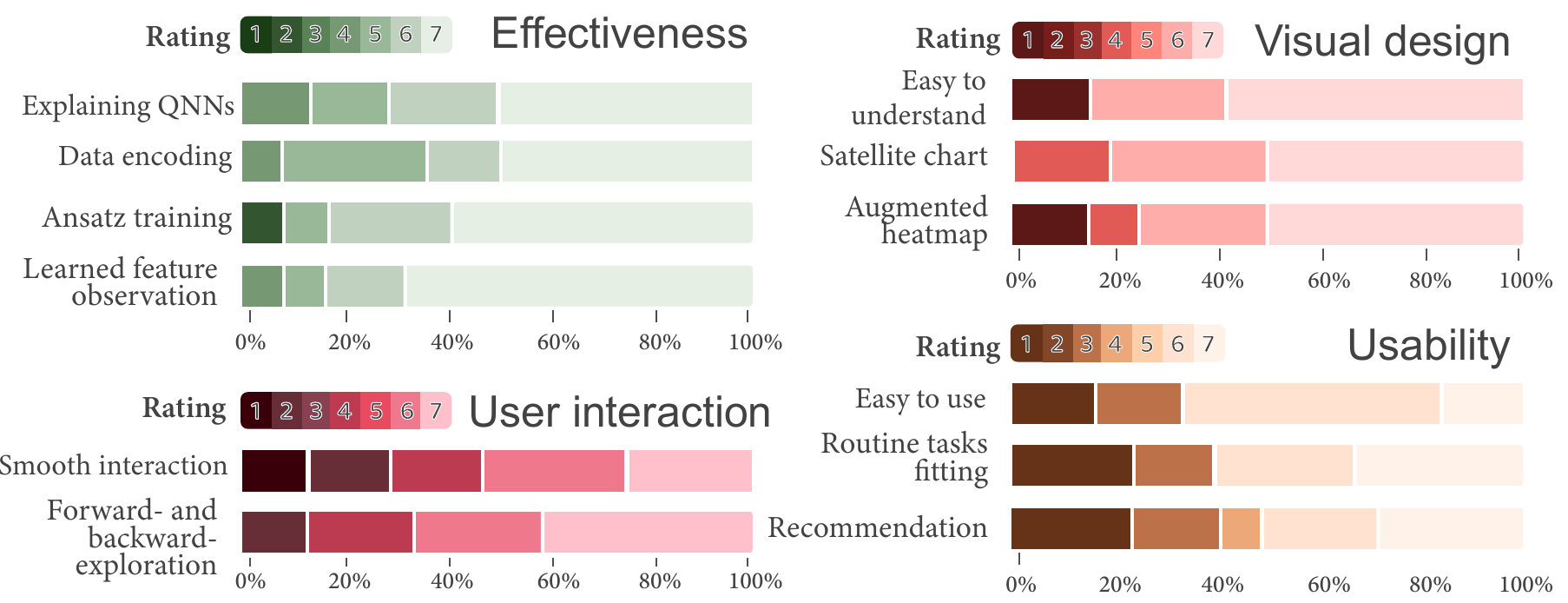}
  \caption{The feedback of the expert interviews. 
}
  \label{fig:5}
\end{figure}

We summarized and reported the qualitative feedback as follows:

\textbf{Effectiveness. }
Most participants ($rating_{mean}=5.81, rating_{sd}=0.83$) praised \toolName's effectiveness in improving the explanability of quantum neural networks.
Specifically, E5 commented ``Through visualization, users can easily observe the impact of each parameter update on the output, providing an efficient means to control and fine-tune hyper-parameters in QML.''
Also, three participants (E2-3, E12) also confirmed the value of multiple views in \toolName{}.
E12 said, ``The first two views (\textit{i.e.,} Encoder View and Ansatz View) are elegant to show how the parameters determine the output, which can provide me with important clues to understand how the network is converged.''
In addition, E3 was really impressed by the Feature View, which can ``intuitively show how the loss is optimized in the certain region of data attributes, making the overall heatmap change across the epochs.'' 
In addition, E11, who studied Quantum Information, also expressed promising usage in other quantum applications such as quantum finance, which is based on quantum neural networks.

\textbf{Visual design. }
Most participants ($rating_{mean}=6.16, rating_{sd}=1.21$) are in favor of the novel designs integrated in \toolName{}.
E7, who works on Quantum Simulator, commented on the satellite chart ``I'm used to using Bloch Sphere to analyze the qubit state when I try to understand the rotation angles in an ansatz. However, to my surprise, \toolName{} can somehow combine the qubit states with basis states. This can directly explain the output which is based on the basis states instead of the qubit one.''
Also, the design principle of the satellite chart gained positive feedback (E6-8) since it uses qubit as the axis to locate the basis states, which can significantly reduce the clutter from $2^N$ to $N$ data entities.
Meanwhile, four participants (E1-3, E8) confirmed that the explanation of the decision boundary is greatly useful, especially in tracing back the basis states of misclassified data points.
E8 also expressed the unique advantage of the coarse- and fine-grained mode of augmented heatmap, which provides a quantum-version viewpoint based on a classical classification heatmap.

\textbf{Usability and user interaction.}
The majority of the participants offered positive feedback regarding the usability  ($rating_{mean}=5.68, rating_{sd}=0.91$) and user interactions  ($rating_{mean}=5.20, rating_{sd}=1.54$).
Notably, three participants (E9-11) praised the implementations of a quantum circuit diagram on the bottom, which can ``give users an intuitive representation of the architecture of the network along with each component (\textit{e.g.,} the encoding step).''
E1-7 confirmed that the system’s interactions are really smooth and easy to use.
Meanwhile, E1 also pointed out that the line charts of loss and accuracy can offer users with a background in classical neural networks a portal to become proficient.
Also, E12 expressed strong will to recommend \toolName{} 
to his colleagues in his research lab.

\section{Discussion}

In this section, we first summarize the lessons we learned during the development of \toolName{}, and then discuss the limitations of \toolName{}.

\subsection{Lessons}

We learned valuable lessons from the system design in collaboration with domain users.

\textbf{Extensible benefits of cross-domain designs. }
According to the process of expert interviews, the indispensable advantages of our proposed designs
have been unequivocally confirmed for the community of variation quantum circuits.
Notably, hinted by the domain expert \textbf{P2} who studies Grover's algorithm, the proposed two designs can also be applied to static quantum circuits and algorithms and further aid their explainability.
For example, satellite chart can allow the detailed analysis of the single-qubit rotations during the inverse Quantum Fourier Transform process of Quantum Phase Estimation (QPE) tasks.
Also, the benefits of the augmented heatmap can be transferred to other research domains with the need to explain the expectation values, like Quantum Simulations and Quantum Error Mitigation.
They believed that the transferable usage of the design could certainly enhance the impact of visualization in the quantum computing community.

\textbf{Strong ties between quantum neural work and classical counterparts. }
By collaborating with domain experts, we realize that it is of great importance to connect the dots between quantum and classical neural networks.
The reason is that developers and practitioners are likely to be familiar with classical machine learning due to its prevalence.
Meanwhile, in accordance with their feedback,  the classical counterparts are more easy-to-understand compared to counter-intuitive quantum information.
Thus, we prefer to use some classical-applicable methods, such as the Evolution View with a heatmap-like design, to flatten the learning curves of understanding quantum neural networks. Also, the statistical charts, like the loss and accuracy line charts, may also mitigate the unfamiliar impression for novice users.

\subsection{Limitations and Future Work}

We introduce the limitations of our current work and future plans.

\textbf{Usage boundary. }
The evaluation has demonstrated that \toolName{} works well for visualizing variational quantum classifiers.
However, with the rapid growth of other directions in quantum machine learning, e.g., quantum generative adversarial networks (qGAN), they are becoming more and more popular.
Specifically, although our visual design, \textit{i.e.,} satellite chart, can support single-qubit state visualization in other types of ansatz,
the tool \toolName{} cannot be directly applied to them for now.
\revise{Thus, we plan to extend the \toolName{} system to enable more diverse directions in quantum machine learning.
Currently, the visualization interface along with the visual design (i.e., satellite chart and augmented heatmap) still has many limitations and can be further enhanced to fit into more complex tasks, such as tuning the hyper-parameters.} 

\textbf{Flexible user interface. }
All the participants appreciated the effectiveness of \toolName{} in enhancing the transparency of quantum neural networks.
However, more actions to improve the flexibility of the interface are recommended to be integrated.
For example, \textbf{P4} commented that the interface could enable users to customize the training datasets by uploading the QASM file.
Also, he also suggested the function of the quantum simulator switching, improving the efficiency of the training process and data extraction. 
In future work, it is worth implementing more interactions to make the user exploration more smooth.

\textbf{Scalability issue. }
The usability of \toolName{} gained positive feedback from all participants.
However, there are still some limitations regarding the scalability of the system.
For example, the representation of original quantum neural networks may suffer from visual clutter when the circuit depth is large.
Also, although we consider utilizing color to encode the probability of qubit states, which can mitigate the scalability issues to some extent.
but the set of circles in the center may still introduce overlapping issues when the basis states are numerous.
To tackle these limitations, we plan to integrate more functionality into the system, such as the grouping of a large amount of data entities.
\section{Conclusion}

In this work, we identified key challenges and requirements
for explaining
the quantum neural networks by working closely with domain experts.
We then introduce \toolName{}, a visualization approach developed for a better understanding of different model components, e.g., encoder, ansatz, and measurement.
We also introduced two model designs, namely a satellite chart and augmented heatmap, to facilitate the detailed analysis of single-qubit states and expectation value measurement, respectively.
We conducted case studies and expert interviews with 12 domain users.
Their ratings and insightful feedback demonstrate the effectiveness and usability of \toolName{}.

\acknowledgments{%
This research was supported by the Lee Kong Chian Fellowship awarded to Yong Wang by Singapore Management
University.
}

\bibliographystyle{abbrv-doi-hyperref}

\bibliography{template}


\end{document}